\documentclass[12pt]{article}
\usepackage[latin9]{inputenc}
\usepackage{color}
\usepackage{amsmath}
\usepackage{esint}
\usepackage[authoryear]{natbib}

\makeatletter

\providecommand{\tabularnewline}{\\}

\@ifundefined{date}{}{\date{}}

\newcommand{\biblist}{\begin{list}{}
{\listparindent 0.0cm \leftmargin 0.50cm \itemindent -0.50 cm
\labelwidth 0 cm \labelsep 0.50 cm
\usecounter{list}}\clubpenalty4000\widowpenalty4000}
\newcommand{\ebiblist}{\end{list}}

\usepackage{latexsym}
\usepackage{multirow}

\newtheorem{theorem}{Theorem}\newtheorem{assumption}{Assumption}\newtheorem{remark}{Remark}

\newcommand{\be}{\begin{equation}}
\newcommand{\en}{\end{equation}}
\newcommand{\bea}{\begin{eqnarray}}
\newcommand{\ena}{\end{eqnarray}}
\newcommand{\ba}{\begin{array}}

\newcommand{\ea}{\end{array}}

\newcommand{\pr}{\mbox{{\rm pr}}}

\newcommand{\T}{\mathrm{\scriptscriptstyle T}}

\newcommand{\var}{ {\mathrm{var}}} 
 
\newcommand{\plim}{ {\mathrm{plim}}} 
\newcommand{\F}{ {\mathcal{F}}} 
\newcommand{\rep}{ {\mathrm{rep}}} 
\newcommand{\pmm}{ {\mathrm{PMM}}} 
 
\newcommand{\nni}{ {\mathrm{NNI}}} 
\newcommand{\HT}{ {\mathrm{HT}}} 
\newcommand{\de}{ {\mathrm{d}}} 
\newcommand{\N}{ {\mathcal{N}}} 
 
\newcommand{\It}{ \mathcal{I}} 
\newtheorem{example}{Example}

\makeatother

\begin{document}
\baselineskip .3in

\title{\textbf{Nearest neighbor imputation for general parameter estimation
in survey sampling}}

\author{Shu Yang\thanks{Department of Statistics, North Carolina State University, North Carolina
27695, U.S.A.}\and Jae Kwang Kim\thanks{Department of Statistics, Iowa State University, Ames, Iowa 50011,
U.S.A.}}
\maketitle
\begin{abstract}
\textcolor{black}{Nearest neighbor imputation is popular for handling
item nonresponse in survey sampling. In this article, we study the
asymptotic properties of the nearest neighbor imputation estimator
for general population parameters, including population means, proportions
and quantiles.} For variance estimation, the conventional bootstrap
inference for matching estimators with fixed number of matches has
been shown to be invalid due to the nonsmoothness nature of the matching
estimator. \textcolor{black}{We propose asymptotically valid replication
variance estimation. The key }strategy is to construct replicates
of the estimator directly based on linear terms, instead of individual
records of variables. A simulation study confirms that the new procedure
provides valid variance estimation. 
\end{abstract}

\newpage{}

\section{Introduction}

Nearest neighbor imputation is popular for handling item nonresponse
in survey sampling. In nearest neighbor imputation, the vector of
the auxiliary variables is directly used in determining the nearest
neighbor. The nearest neighbor is then used as a donor for hot deck
imputation. Although these imputation methods have a long history
of application, there are relatively few papers on investigating their
asymptotic properties. \citet{sande1979personal} discussed nearest
neighbor rules in statistical estimation with hot-deck imputation.
\citet{lee1994experiments} studied methods of nearest neighbor imputation.
\citet{chen2000nearest,chen2001jackknife} have developed a nice set
of asymptotic theories for the nearest neighbor imputation estimator.
\citet{abadie2006large} studied the matching estimator to estimate
the average treatment effect from observational studies. \citet{shao2008confidence}
proposed methods for constructing confidence intervals for population
means and quantiles with nearest neighbor imputation. \citet{kim2011variance}
presented an application of nearest neighbor imputation for the US
Census long form data. However, most of these studies discussed either
with a $1$-dimensional covariate or only for mean estimation, which
is restrictive both theoretically and practically. 

Survey statisticians are often interested in various finite population
quantities, such as the population means, proportions and quantiles
\citep{francisco1991quantile,wu2001model,berger2003variance}, to
name a few. Some corresponding sample estimators should be treated
differently than others. For example, estimators of population quantiles
involve nondifferentiable functions of estimated quantities. Moreover,
there often are more than one auxiliary covariates available to facilitate
nearest neighbor imputation. The current framework of nearest neighbor
imputation can not cover inferences in these settings. 

In this article, we provide a framework of nearest neighbor imputation
for general parameter estimation in survey sampling. In general, the
matching estimators are not root-$n$ consistent \citep{abadie2006large},
where $n$ is the sample size. Based on a scalar matching variable
$m$ summarizing all auxiliary information, we show that nearest neighbor
imputation can provide consistent estimators for a fairly general
class of parameters. If the matching variable is chosen to be the
mean function of the study variable, our method resembles prediction
mean matching imputation. However, the validity of predictive mean
matching requires the mean function to be correctly specified. Here,
we show that the consistency of the nearest neighbor imputation estimator
only requires the matching variable to satisfy certain Lipschitz continuity
condition. For inference, intrinsically the nearest neighbor imputation
estimator with fixed number of matches is not smooth. The lack of
smoothness makes the conventional replication methods invalid for
variance estimation, mainly because the naive replication method distorts
the distribution of the number of times each unit is used as a match.
We propose new replication variance estimation. Based on the linear
representation of the nearest neighbor imputation estimator, we construct
replicates of the estimator directly based on its linear terms. In
this way, the distribution of the number of times each unit is used
as a match can be preserved, which leads to a valid variance estimation.
Furthermore, our replication variance method is flexible, which can
accommodate bootstrap and jackknife, among others.

\section{Basic Setup}

Let $\mathcal{F}_{N}=\{(x_{i},y_{i},\delta_{i}):i=1,\ldots,N\}$ denote
a finite population, where $x_{i}$ is a $p$-dimensional vector of
covariates, which is always observed, $y_{i}$ has missing values,
and $\delta_{i}$ is the response indicator of $y_{i}$, i.e., $\delta_{i}=1$
if $y_{i}$ is observed and $0$ if it is missing. The $\delta_{i}$'s
are defined throughout the finite population, as in Fay (1992), \citet{shao1999variance},
and \citet{kim2006replication}. We assume that $\F_{N}$ is a random
sample from a superpopulation model $\zeta$, and $N$ is known. Our
objective is to estimate the finite population parameter defined through
$\mu_{g}=N^{-1}\sum_{i=1}^{N}g(y_{i})$ for some known $g(\cdot)$,
or $\xi_{N}=\inf\{\xi:S_{N}(\xi)\geq0\}$, where $S_{N}(\xi)=N^{-1}\sum_{i=1}^{N}s(y_{i}-\xi)$
and $s(\cdot)$ is a univariate real function. These parameters are
fairly general, which cover many parameters of interest in survey
sampling. For example, let $g(y)=y$, $\mu_{g}$ is the population
mean of $y$, $N^{-1}\sum_{i=1}^{N}y_{i}$. Let $g(y)=I(y<c)$ for
some constant $c$, $\mu_{g}$ is the population proportion of $y$
less than $c$, $N^{-1}\sum_{i=1}^{N}I(y_{i}<c)$. Let $s(y_{i}-\xi)=I(y_{i}\leq\xi)-\alpha$,
$\xi_{N}$ is the population $\alpha$th quantile.

Let $A$ denote an index set of the sample selected by a probability
sampling design. Let $I_{i}$ be the sampling indicator function,
i.e., $I_{i}=1$ if unit $i$ is selected into the sample, and $I_{i}=0$
otherwise. Suppose that $\pi_{i}$, the first-order inclusion probability
of unit $i$, is positive and known throughout the sample. If $y_{i}$
were fully observed throughout the sample, the sample estimator of
$\mu_{g}$ and $\xi_{N}$ are $\hat{\mu}_{g}=N^{-1}\sum_{i\in A}\pi_{i}^{-1}g(y_{i})$
and $\hat{\xi}=\inf\{\xi:\hat{S}_{N}(\xi)\geq0\}$ with $\hat{S}_{N}(\xi)=N^{-1}\sum_{i\in A}\pi_{i}^{-1}s(y_{i}-\xi)$,
respectively. 

We make the following assumption for the missing data process.

\begin{assumption}[Missing at random and positivity]\label{asmp:MAR}The
missing data process satisfies $\pr(\delta=1\mid x,y)=\pr(\delta=1\mid x)$,
which is denoted by $p(x)$, and with probability $1$, $p(x)>\epsilon$
for a constant $\epsilon>0$.

\end{assumption}

Our primary focus will be on the imputation estimators of $\mu_{g}$
and $\xi_{N}$ given by $\hat{\mu}_{g,I}=N^{-1}\sum_{i\in A}\pi_{i}^{-1}\left\{ \delta_{i}g(y_{i})+(1-\delta_{i})g(y_{i}^{*})\right\} $
and $\hat{\xi}_{I}=\inf\{\xi:\hat{S}_{I}(\xi)\geq0\}$, with $\hat{S}_{I}(\xi)=\hat{N}^{-1}\sum_{i\in A}\pi_{i}^{-1}s(y_{i}-\xi)\left\{ \delta_{i}s(y_{i}-\xi)+(1-\delta_{i})s(y_{i}^{*}-\xi)\right\} $,
where $y_{i}^{*}$ is an imputed value of $y_{i}$ for unit $i$ with
$\delta_{i}=0$. To find suitable imputed values, the classical nearest
neighbor imputation can be described in the following steps: 
\begin{description}
\item [{Step$\ $1.}] For each unit $i$ with $\delta_{i}=0$, find the
nearest neighbor from the respondents with the minimum distance between
$x_{j}$ and $x_{i}$. Let $i(1)$ be the index set of its nearest
neighbor, which satisfies $d(x_{i(1)},x_{i})\le d(x_{j},x_{i}),$
for $j\in A_{R}$, where $d(x_{i},x_{j})$ is a distance function
between $x_{i}$ and $x_{j}$. For example, $d(x_{i},x_{j})=||x_{i}-x_{j}||$,
where $||x||=(x^{\T}x)^{1/2}$.\textcolor{black}{{} Other norms of the
form $||x||_{D}=(x^{\T}Dx)^{1/2}$, where $D$ is a positive definite
symmetric matrix $D$, are equivalent to the Euclidean norm, since
$||x||_{D}=\{(Qx)^{\T}(Qx)\}^{1/2}=||Qx||$ with $Q^{\T}Q=D$. In
particular, Mahalanobis distance is commonly used, where $D=\hat{\Sigma}^{-1}$
with $\hat{\Sigma}$ the empirical covariance matrix of $x$. } 
\item [{Step$\ $2.}] The nearest neighbor imputation estimators of $\mu_{g}$
and $\xi_{N}$ are computed by 
\begin{equation}
\hat{\mu}_{g,\nni}=\frac{1}{N}\sum_{i\in A}\frac{1}{\pi_{i}}\left\{ \delta_{i}g(y_{i})+(1-\delta_{i})g(y_{i(1)})\right\} ,\label{eq:nni}
\end{equation}
and $\hat{\xi}_{\nni}=\inf\{\xi:\hat{S}_{\nni}(\xi)\geq0\}$, respectively,
with 
\begin{equation}
\hat{S}_{\nni}(\xi)=\frac{1}{N}\sum_{i\in A}\pi_{i}^{-1}\left\{ \delta_{i}s(y_{i}-\xi)+(1-\delta_{i})s(y_{i(1)}-\xi)\right\} .\label{eq:nni_s}
\end{equation}
\end{description}
In (\ref{eq:nni}) and (\ref{eq:nni_s}), the imputed values are real
observations. 

\section{Main result}

For asymptotic inference, we follow the framework of \citet{isaki1982survey}
where the asymptotic properties of estimators are established under
a fixed sequence of populations and a corresponding sequence of random
samples. Denote $E_{p}(\cdot)$ and $\var_{p}(\cdot)$ to be the expectation
and the variance under the sampling design, respectively. We impose
the following regularity conditions on the sampling design.

\begin{assumption}\label{asmp:sampling} (i) There exist positive
constants $C_{1}$ and $C_{2}$ such that $C_{1}\le\pi_{i}Nn^{-1}\le C_{2},$
for $i=1,\ldots,N$; (ii) the sequence of the Hotvitz-Thompson estimators
$\hat{\mu}_{g,\HT}=N^{-1}\sum_{i\in A}\pi_{i}^{-1}g(y_{i})$ satisfies
$\var_{p}(\hat{\mu}_{g,\HT})=O(n^{-1})$ and $\{\var_{p}(\hat{\mu}_{g,\HT})\}^{-1/2}(\hat{\mu}_{g,\HT}-\mu_{g})\mid\mathcal{F}_{N}\rightarrow\N(0,1)$
in distribution, as $n\rightarrow\infty$.

\end{assumption}

Assumption \ref{asmp:sampling} is a widely accepted assumption in
survey sampling \citep{fuller2009sampling}.

We introduce additional notation. Let $A=A_{R}\cup A_{M}$, where
$A_{R}$ and $A_{M}$ are the sets of respondents and nonrespondents,
respectively. Define $d_{ij}=1$ if $y_{j(1)}=y_{i}$, i.e., unit
$i$ is used as a donor for unit $j\in A_{M}$, and $d_{ij}=0$ otherwise.
We write $\hat{\mu}_{g,\nni}$ in (\ref{eq:nni}) as 
\begin{equation}
\hat{\mu}_{g,\nni}=\frac{1}{N}\left\{ \sum_{i\in A}\frac{1}{\pi_{i}}\delta_{i}g(y_{i})+\sum_{j\in A}\frac{1-\delta_{j}}{\pi_{j}}\sum_{i\in A}\delta_{i}d_{ij}g(y_{i})\right\} =\frac{1}{N}\sum_{i\in A}\frac{\delta_{i}}{\pi_{i}}(1+k_{i})g(y_{i}),\label{eq:expression}
\end{equation}
with 
\begin{equation}
k_{i}=\sum_{j\in A}\frac{\pi_{i}}{\pi_{j}}(1-\delta_{j})d_{ij}.\label{eq:ki}
\end{equation}
Under simple random sampling, $k_{i}=\sum_{j\in A}(1-\delta_{j})d_{ij}$
is the number of times that unit $i$ is used as the nearest neighbor
for nonrespondents. 

To study the asymptotic properties of the nearest neighbor imputation
estimator $\hat{\mu}_{g,\nni}$, we use the following decomposition:
\begin{equation}
n^{1/2}(\hat{\mu}_{g,\nni}-\mu_{g})=D_{N}+B_{N},\label{eq:decomposition}
\end{equation}
 where 
\begin{equation}
D_{N}=n^{1/2}\left[\frac{1}{N}\sum_{i\in A}\frac{1}{\pi_{i}}\left\{ \mu_{g}(x_{i})+\delta_{i}(1+k_{i})\{g(y_{i})-\mu_{g}(x_{i})\right\} -\mu_{g}\right],\label{eq:DN}
\end{equation}
and 
\begin{equation}
B_{N}=\frac{n^{1/2}}{N}\sum_{i\in A}\frac{1}{\pi_{i}}(1-\delta_{i})\{\mu_{g}(x_{i(1)})-\mu_{g}(x_{i})\}.\label{eq:bias}
\end{equation}
The difference $\mu_{g}(x_{i(1)})-\mu_{g}(x_{i})$ accounts for the
matching discrepancy, and $B_{N}$ contributes to the asymptotic bias
of the matching estimator. In general, if $x$ is $p$-dimensional,
\citet{abadie2006large} showed that $d(x_{i(1)},x_{i})=O_{p}(n^{-1/p})$.
Therefore, for nearest neighbor imputation with $p\geq2$, the bias
$B_{N}=O_{p}(n^{1/2-1/p})\neq o_{p}(1)$ is not negligible. 

To address for the matching discrepancy due to a non-scalar $x$,
we first summarize the covariate information into a scalar matching
variable $m=m(x)$, and then apply nearest neighbor imputation based
on this scalar variable. For simplicity of notation, we may suppress
the dependence of $m$ on $x$ if there is no ambiguity. For nearest
neighbor imputation with a scalar matching variable, we then have
$p=1$ and $B_{N}=O_{p}(n^{-1/2})=o_{p}(1)$. We assume the superpopulation
model and the matching variable $m$ satisfy the following assumption.

\begin{assumption}\label{asmp:m} (i) The matching variable $m$
has a compact and convex support, with density bounded and bounded
away from zero. Let $f_{1}(m)$ and $f_{0}(m)$ be the conditional
density of $m$ given $\delta=1$ and $\delta=0$, respectively. Suppose
that there exist constants $C_{1L}$ and $C_{1U}$ such that $C_{1L}\leq f_{1}(m)/f_{0}(m)\leq C_{1U}$;
(ii) $\mu_{g}(x)=E\{g(y)\mid x\}$ and $\mu_{s}(\xi,x)=E\{s(y-\xi)\mid x\}$
sastisfy certain Lipschitz continuous condition; i.e., there exists
a constant $C_{2}$ such that $|\mu_{g}(x_{i})-\mu_{g}(x_{j})|<C_{2}|m_{i}-m_{j}|$
and $|\mu_{s}(\xi,x_{i})-\mu_{s}(\xi,x_{j})|<C_{2}|m_{i}-m_{j}|$
for any $i$ and $j$; (iii) there exists $\delta>0$ such that $E(|g(y)|^{2+\delta}\mid x)$
and $E(|s(y-\xi)|^{2+\delta}\mid x)$ are uniformly bounded for any
$x$ and $\xi$ in the neighborhood of $\xi_{N}$. 

\end{assumption}

Assumption \ref{asmp:m} (i) a convenient regularity condition \citep{abadie2006large}.
Assumption \ref{asmp:m} (ii) imposes a smoothness condition for $\mu_{g}(x)$,
$\mu_{s}(\xi,x)$ and $m(x)$, which is not restrictive \citep{chen2000nearest}.
Assumption \ref{asmp:m} (iii) is a moment condition for establishing
the central limit theorem.

We establish the asymptotic distribution of $\hat{\mu}_{g,\nni}$,
with the proof deferred to the Appendix. 

\begin{theorem}\label{Thm:1}Under Assumptions \ref{asmp:MAR}\textendash \ref{asmp:sampling},
suppose that $\mu_{g}(x)=E\{g(y)\mid x\}$ and $\sigma_{g}^{2}(x)=\mathrm{var}\{g(y)\mid x\}$.
Then, $n^{1/2}\{\hat{\mu}_{g,\nni}-\mu_{g}\}\rightarrow\N(0,V_{g})$
in distribution, as $n\rightarrow\infty$, where 
\begin{equation}
V_{g}=V_{g}^{\mu}+V_{g}^{e}\label{eq:V1}
\end{equation}
with $V_{g}^{\mu}=\lim_{n\rightarrow\infty}nN^{-2}E[\var_{p}\{\sum_{i\in A}\pi_{i}^{-1}\mu_{g}(x_{i})\}],$
$V_{g}^{e}=\lim_{n\rightarrow\infty}nN^{-2}E[\sum_{i\in A}\{\pi_{i}^{-1}\delta_{i}(1+k_{i})-1\}^{2}\sigma_{g}^{2}(x_{i})],$
and $k_{i}$ is defined in (\ref{eq:ki}).

\end{theorem}

We now establish a similar result for $\hat{\xi}_{\nni}$, with the
proof deferred to the Appendix. 

\begin{theorem}\label{Thm:2}Under Assumptions \ref{asmp:MAR}\textendash \ref{asmp:sampling},
suppose the population parameter $\xi_{N}$ and the population estimating
function $S_{N}(\cdot)$ satisfy certain regularity conditions specified
in Assumptions \ref{asmp:sN} and \ref{asmp:sN-1}. We obtain an asymptotic
linearization representation of $\hat{\xi}_{\nni}$: 
\begin{equation}
n^{1/2}(\hat{\xi}_{\nni}-\xi)=-n^{1/2}\{\hat{S}_{\nni}(\xi)-S_{N}(\xi)\}/S'(\xi)+o_{p}(1).\label{eq:s2}
\end{equation}
It follows that $n^{1/2}(\hat{\xi}_{\nni}-\xi_{N})\rightarrow\N(0,V_{\xi})$
in distribution, as $n\rightarrow\infty$, where $V_{\xi}=\dot{S}(\xi_{N})^{-2}\var\{\hat{S}_{\nni}(\xi_{N})\}$,
$\dot{S}(\xi_{N})=\de S(\xi_{N})/\de\xi$, and
\begin{multline}
\var\{\hat{S}_{\nni}(\xi_{N})\}=\lim_{n\rightarrow\infty}\frac{n}{N^{2}}E\left(\var_{p}\left[\sum_{i\in A}\frac{E\{s(y_{i}-\xi_{N})\mid x_{i}\}}{\pi_{i}}\right]\right)\\
+\plim\frac{n}{N^{2}}\sum_{i=1}^{N}\left\{ \frac{I_{i}}{\pi_{i}}\delta_{i}(1+k_{i})-1\right\} ^{2}\var\left[s(y_{i}-\xi_{N})-E\{s(y_{i}-\xi_{N})\mid x_{i}\}\mid x_{i}\right],\label{eq:VS}
\end{multline}
and $k_{i}$ is defined in (\ref{eq:ki}).

\end{theorem}

For illustration, we use quantile estimation as an example. 

\begin{example}[Quantile estimation]\label{eg quantile estimation }The
estimating function for the $\alpha$th quantile is $s(y_{i}-\xi)=I(y_{i}\leq\xi)-\alpha$,
and the population estimating equation $S_{\alpha,N}(\xi)=F_{N}(\xi)-\alpha$,
where $F_{N}(\xi)=N^{-1}\sum_{i=1}^{N}I(y_{i}\leq\xi)$. The nearest
neighbor imputation estimator $\hat{\xi}_{\alpha,\nni}$ is defined
as 
\[
\hat{\xi}_{\alpha,\nni}=\inf\{\xi:\hat{S}_{\alpha,\nni}(\xi)\geq0\},
\]
where $\hat{S}_{\alpha,\nni}(\xi)=\hat{F}_{\nni}(\xi)-\alpha$, $\hat{F}_{\nni}(\xi)=\hat{N}^{-1}\sum_{i\in A}\pi_{i}^{-1}\delta_{i}(1+k_{i})I(y_{i}\leq\xi)$,
$\hat{N}=\sum_{i\in A}\pi_{i}^{-1}$, and $k_{i}$ is defined in (\ref{eq:ki}).
Let $F(\xi)=\text{\pr}(y\leq\xi)$ be the cumulative distribution
function of $y$. Then, $\hat{F}_{\nni}(\xi)$ is a Hajek estimator
for $F(\xi)$, which is asymptotically equivalent to the one using
$N$ instead of $\hat{N}$. Even with a known $N$, it is necessary
to use $\hat{N}$ because $\hat{F}_{\nni}(\xi)$ for $\xi=\infty$
should be $1$. The limiting function of $S_{\alpha,N}(\xi)$ is $S_{\alpha}(\xi)=F(\xi)-\alpha$.
The asymptotic linearization representation of $\hat{\xi}_{\alpha,\nni}$
is 
\begin{equation}
\hat{\xi}_{\alpha,\nni}-\xi=-\frac{\hat{F}_{\nni}(\xi)-F_{N}(\xi)}{f(\xi)}+o_{p}(n^{-1/2}),\label{eq:quantile nni}
\end{equation}
where $f(\xi)=F'(\xi)$. Expression (\ref{eq:quantile nni}) is called
the Bahadur-type representation for $\hat{\xi}_{\alpha,\nni}$ \citep{francisco1991quantile}. 

\end{example}

\begin{remark}[The choice of the scalar matching variable]

By judicious choice, the scalar matching variable should ensure that
Assumption \ref{asmp:m} holds. If the conditional mean function of
the outcome variable given the covariates is feasible, we can choose
the matching variable to be the conditional mean function. We note
that in this case the proposed nearest neighbor imputation resembles
predictive mean matching imputation. However, our method is more general
than predictive mean matching imputation, because the latter requires
the mean function to be correctly specified. 

\end{remark}

\section{Replication variance estimation\label{sec:Replication-variance-estimation} }

We consider replication variance estimation \citep{rust1996variance,wolter2007introduction}
for nearest neighbor imputation. 

Let $\hat{\mu}_{g}$ be the Horvitz-Thompson estimator of $\mu_{g}.$
The replication variance estimator of $\hat{\mu}_{g}$ takes the form
of 
\begin{equation}
\hat{V}_{\rep}(\hat{\mu}_{g})=\sum_{k=1}^{L}c_{k}(\hat{\mu}_{g}^{(k)}-\hat{\mu}_{g})^{2},\label{eq:replication variance}
\end{equation}
where $L$ is the number of replicates, $c_{k}$ is the $k$th replication
factor, and $\hat{\mu}_{g}^{(k)}$ is the $k$th replicate of $\hat{\mu}_{g}$.
When $\hat{\mu}_{g}=\sum_{i\in A}\omega_{i}g(y_{i})$, we can write
the replicate of $\hat{\mu}_{g}$ as $\hat{\mu}_{g}^{(k)}=\sum_{i\in A}\omega_{i}^{(k)}g(y_{i})$
with some $\omega_{i}^{(k)}$ for $i\in A$. The replications are
constructed such that $E_{p}\{\hat{V}_{\rep}(\hat{\mu}_{g})\}=\var_{p}(\hat{\mu}_{g})\{1+o(1)\}$.

We propose a new replication variance estimation for $\hat{\mu}_{g,\nni}$.
Let $\omega_{i}=N^{-1}\pi_{i}^{-1}$. Write $\hat{\mu}_{g,\nni}-\mu_{g}=(\hat{\mu}_{g,\pmm}-\hat{\psi}_{\HT})+(\hat{\psi}_{\HT}-\mu_{\psi})+(\mu_{\psi}-\mu_{g}),$
where $\hat{\psi}_{\HT}=\sum_{i\in A}\omega_{i}\psi_{i}$, $\psi_{i}=\mu_{g}(x_{i})+\delta_{i}(1+k_{i})\{g(y_{i})-\mu_{g}(x_{i})\}$,
$\mu_{\psi}=N^{-1}\sum_{i=1}^{N}\psi_{i}$. Because $\mu_{g,\nni}-\hat{\psi}_{\HT}=o_{p}(n^{-1/2})$
by Theorem \ref{Thm:1} and $\mu_{\psi}-\mu_{g}=O_{p}(N^{-1/2})$,
we have $\hat{\mu}_{g,\nni}-\mu_{g}=\hat{\psi}_{\HT}-\mu_{\psi}+o_{p}(n^{-1/2})$,
if $nN^{-1}=o(1)$. Therefore, with negligible sampling fractions,
it is sufficient to estimate the variance of $\hat{\psi}_{\HT}-\mu_{\psi}$.
Because $E(\hat{\psi}_{\HT}-\mu_{\psi}\mid\mathcal{F}_{N})=0$, we
have $\var(\hat{\psi}_{\HT}-\mu_{\psi})=E\{\var(\hat{\psi}_{\HT}-\mu_{\psi}\mid\mathcal{F}_{N})\},$
which is essentially the sampling variance of $\hat{\psi}_{\HT}$.
This suggests that we can treat $\{\psi_{i}:i\in A\}$ as pseudo observations
in applying replication variance estimator. \citet{otsu2015bootstrap}
used a similar idea to develop a wild bootstrap technique for a matching
estimator. To be specific, we construct replicates of $\hat{\psi}_{\HT}$
as follows: $\hat{\psi}_{\HT}^{(k)}=\sum_{i\in A}\omega_{i}^{(k)}\psi_{i},$
where $\omega_{i}^{(k)}$ is the replication weight that account for
complex sampling design. The replication variance estimator of $\hat{\psi}_{\HT}$
is obtained by applying $\hat{V}_{\rep}(\cdot)$ in (\ref{eq:replication variance})
for the above replicates $\hat{\psi}_{\HT}^{(k)}$. It follows that
$E\{\hat{V}_{\rep}(\hat{\psi}_{\HT})\}=\var(\hat{\psi}_{\HT}-\mu_{\psi})\{1+o(1)\}=\var(\hat{\mu}_{g,\nni}-\mu_{g})\{1+o(1)\}$.
Because $\mu_{g}(x)$ is unknown, we use a plug-in kernel estimator
$\hat{\mu}_{g}(x)$. 

\textcolor{black}{In summary, the new replication variance estimation
for $\hat{\mu}_{g,\nni}$ proceeds as follows:}
\begin{description}
\item [{Step$\ 1.$}] Obtain a consistent kernel estimator $\hat{\mu}_{g}(x)$. 
\item [{Step$\ 2.$}] \textcolor{black}{Construct replicates of $\hat{\mu}_{g,\nni}$
as
\begin{eqnarray}
\hat{\mu}_{g,\nni}^{(k)} & = & \sum_{i\in A}\omega_{i}^{(k)}[\hat{\mu}_{g}(x_{i})+\delta_{i}(1+k_{i})\{g(y_{i})-\hat{\mu}_{g}(x_{i})\}],\label{eq:k-rep-nni}
\end{eqnarray}
where $\omega_{i}^{(k)}$ is the $k$th replication weight for unit
$i$. }
\item [{Step$\ 3.$}] \textcolor{black}{Apply $\hat{V}_{\rep}(\cdot)$
in (\ref{eq:replication variance}) for the above replicates to obtain
the replication variance estimator of $\hat{\mu}_{g,\nni}$.}
\end{description}
We now consider a replication variance estimation for $\hat{\xi}_{\nni}$.
Following the previous section, we directly obtain the asymptotic
variance of $\hat{\xi}_{\nni}$ using $\var\{\hat{S}_{\nni}(\xi)\}$
and $S'(\xi)$. First to estimate $\var\{\hat{S}_{\nni}(\xi)\}$,
we can use the similar replication variance estimation earlier in
this section. \textcolor{black}{Now to estimate $S'(\xi)$, we follow
the kernel-based derivative estimation of \citet{deville1999variance}:
\begin{equation}
\hat{S}'(\xi)=\frac{1}{Nh}\sum_{i\in A}\frac{1}{\pi_{i}}\int s(y_{i}-x)K'\left(\frac{\xi-x}{h}\right)\de x\label{eq:derivativeS}
\end{equation}
where $K(\cdot)$ is a kernel function in $\mathcal{R}$, $K'(x)=\de K(x)/\de x$,
and $h$ is the bandwidth. }Under Assumption \ref{asump:10} for the
kernel function and bandwidth and previously stated regularity conditions
on the superpopulations and sampling designs, the kernel-based estimator
(\ref{eq:derivativeS}) is consistent for $S'(\xi)$.

\textcolor{black}{In summary, the new replication variance estimation
for $\hat{\xi}_{\nni}$ proceeds as follows:}
\begin{description}
\item [{Step$\ 1.$}] Obtain a consistent kernel estimator $\hat{\mu}_{s}(\hat{\xi}_{\nni},x)$
\item [{Step$\ 2.$}] \textcolor{black}{Construct replicates }of $\hat{S}_{\nni}(\hat{\xi}_{\nni})$\textcolor{black}{{}
as
\begin{equation}
\hat{S}_{\nni}^{(k)}(\hat{\xi}_{\nni})=\sum_{i\in A}\omega_{i}^{(k)}[\hat{\mu}_{s}(\hat{\xi}_{\nni},x_{i})+\delta_{i}(1+k_{i})\{s(y_{i}-\hat{\xi}_{\nni})-\hat{\mu}_{s}(\hat{\xi}_{\nni},x_{i})\}].\label{eq:k-rep-nni-1}
\end{equation}
}
\item [{Step$\ 3.$}] \textcolor{black}{Apply $\hat{V}_{\rep}(\cdot)$
in (\ref{eq:replication variance}) for the above replicates to obtain
the replication variance estimator of $\hat{S}_{\nni}(\hat{\xi}_{\nni})$,
denoted as }$\hat{V}_{\rep}\{\hat{S}_{\nni}(\hat{\xi}_{\nni})\}.$
\item [{Step$\ 4.$}] Obtain the kernel-based derivative estimator $\hat{S}'(\hat{\xi}_{\nni})$,
where $\hat{S}'(\xi)$ is defined in (\ref{eq:derivativeS}). 
\item [{Step$\ 5.$}] Calculate the variance estimator of $\hat{\xi}_{\nni}$
as $\hat{S}'(\hat{\xi}_{\nni})^{-2}\hat{V}_{\rep}\{\hat{S}_{\nni}(\hat{\xi}_{\nni})\}$. 
\end{description}
For illustration, we continue with Example \ref{eg quantile estimation }.

\begin{example}[Quantile estimation (Cont.)]

Obtain kernel-based estimators for $F(\xi)=\pr(y\leq\xi)$ and $f(\xi)$,
denoted as $\hat{F}(\xi)$ and $\hat{f}(\xi)$, respectively. Construct
replicates of $\hat{F}_{\nni}(\hat{\xi}_{\alpha,\nni})$\textcolor{black}{{}
as $\hat{F}_{\nni}^{(k)}(\hat{\xi}_{\alpha,\nni})=\sum_{i\in A}\omega_{i}^{(k)}[\hat{F}(\hat{\xi}_{\alpha,\nni})+\delta_{i}(1+k_{i})\{I(y_{i}\leq\hat{\xi}_{\alpha,\nni})-\hat{F}(\hat{\xi}_{\alpha,\nni})\}].$
Apply $\hat{V}_{\rep}(\cdot)$ in (\ref{eq:replication variance})
for the above replicates to obtain the replication variance estimator
of $\hat{F}_{\nni}(\hat{\xi}_{\alpha,\nni})$, denoted as }$\hat{V}_{\rep}\{\hat{F}_{\nni}(\hat{\xi}_{\alpha,\nni})\}.$
Calculate the variance estimator of $\hat{\xi}_{\alpha,\nni}$ as
$\hat{f}(\hat{\xi}_{\alpha,\nni})^{-2}\hat{V}_{\rep}\{\hat{F}_{\nni}(\hat{\xi}_{\alpha,\nni})\}$. 

\end{example}

\begin{theorem}\label{Thm:ve1}Under the assumptions in Theorem \ref{Thm:2},
suppose that $\hat{V}_{\rep}(\hat{\mu}_{g})$ in (\ref{eq:replication variance})
is consistent for $\var_{p}(\hat{\mu}_{g})$. Then, if $nN^{-1}=o(1)$,
the replication variance estimators for $\hat{\mu}_{g,\nni}$ is consistent,
i.e., $n\hat{V}_{\rep}\{\hat{\mu}_{g,\nni}\}/V_{g}\rightarrow1$ in
probability, as $n\rightarrow\infty$, where $\hat{V}_{\rep}(\cdot)$
is given in (\ref{eq:replication variance}), the replicates of $\hat{\mu}_{g,\nni}$
are given in (\ref{eq:k-rep-nni}), and $V_{g}$ is given in (\ref{eq:V1}).

Given that the kernel-based estimator $\hat{S}'(\xi)$ in (\ref{eq:derivativeS})
is consistent for $S'(\xi)$, the replication variance estimators
for $\hat{\xi}_{\nni}$ is consistent, i.e., $n\hat{V}_{\rep}\{\hat{\xi}_{\nni}\}/V_{\xi}\rightarrow1$
in probability, as $n\rightarrow\infty$, where $\hat{V}_{\rep}(\cdot)$
is given in (\ref{eq:replication variance}), the replicates of $\hat{S}_{\nni}^{(k)}(\hat{\xi}_{\nni})$
are given in (\ref{eq:k-rep-nni-1}), and $V_{\xi}$ is given in (\ref{eq:VS}).

\end{theorem}

The formal proof follows by straightforward asymptotic bounding arguments
from the assumptions and therefore is omitted. 

\section{Simulation study}

In this simulation study, we investigate the performance of the proposed
replication variance estimation. For generating finite populations
of size $N=50,000$: first, let $x_{1i}$, $x_{2i}$ and $x_{3i}$
be generated independently from Uniform$[0,1]$, and $x_{4i}$, $x_{5i}$
and $x_{6i}$ and $e_{i}$ be generated independently from $\N(0,1)$;
then, let $y_{i}$ be generated as (P1) $y_{i}=-1+x_{1i}+x_{2i}+e_{i}$,
(P2) $y_{i}=-1.5+x_{1i}+x_{2i}+x_{3i}+x_{4i}+e_{i}$, (P3) $y_{i}=-1.5+x_{1i}+\cdots+x_{6i}+e_{i}$,
(P4) $y_{i}=-1+x_{1i}+x_{2i}+x_{1i}^{2}+x_{2i}^{2}-2/3+e_{i}$, (P5)
$y_{i}=-1.5+x_{1i}+x_{2i}+x_{3i}+x_{4i}+x_{1i}^{2}+x_{2i}^{2}-2/3+e_{i}$
and (P6) $y_{i}=-1.5+x_{1i}+\cdots+x_{6i}+x_{1i}^{2}+x_{2i}^{2}-2/3+e_{i}$.
The covariates are fully observed, but $y_{i}$ is not. The response
indicator of $y_{i}$, $\delta_{i}$, is generated from Bernoulli$(p_{i})$
with logit\{$p(x_{i})\}=x_{i}^{\T}1$, where $x_{i}$ includes all
corresponding covariates under each data generating mechanism and
$1$ is a vector of $1$ with a compatible length. This results in
the average response rate about $75\%$. The parameters of interest
are $\mu=N^{-1}\sum_{i=1}^{N}y_{i}$, $\eta=N^{-1}\sum_{i=1}^{N}I(y_{i}<c)$,
where $c$ is the $80$th quantile such that the true value of $\eta$
is $0.8$, and the median $\xi$. To generate samples, we consider
two sampling designs: (S1) simple random sampling with $n=800$; (S2)
probability proportional to size sampling. In (S2), for each unit
in the population, we generate a size variable $s_{i}$ as $\log(|y_{i}+\nu_{i}|+4)$,
where $\nu_{i}\sim\N(0,1)$. The selection probability is specified
as $\pi_{i}=400s_{i}/\sum_{i=1}^{N}s_{i}$. Therefore, (S2) is informative,
where units with larger $y_{i}$ values have larger probabilities
to be selected into the sample.

For nearest neighbor imputation, the matching scalar variable $m$
is set to be the conditional mean function of $y$ given $x$, $m(x)$,
approximated by power series estimation. For investigating the effect
of the matching variable, we consider the power series including all
first and second order terms under (P1)\textendash (P3) and only first
order terms under (P4)\textendash (P6), so that $m(x)$ is accurate
for the mean function under (P1)\textendash (P3) but inaccurate under
(P4)\textendash (P6). We construct $95\%$ confidence intervals using
$(\hat{\mu}_{I}-z_{0.975}\hat{V}_{I}^{1/2},\hat{\mu}_{I}+z_{0.975}\hat{V}_{I}^{1/2})$,
where $\hat{\mu}_{I}$ is the joint estimate and $\hat{V}_{I}$ is
the variance estimate obtained by the proposed jackknife variance
estimation and a naive jackknife variance estimation that calculates
a sample estimator for each replicate. For the jackknife replication
method under (S2), in the $k$th replicate, the replication weights
are $\omega_{i}^{*(k)}=n\omega_{i}/(n-1)$ for all $i\neq k$, and
$\omega_{k}^{*(k)}=0$. In the proposed jackknife variance estimation,
the $k$th replicates of $\hat{\mu}_{\nni}$, $\hat{\eta}_{\nni}$
and $\hat{\xi}_{\nni}$ are given by 
\[
\hat{\mu}_{\nni}^{(k)}=\sum_{i=1}^{n}\omega_{i}^{(k)}[\hat{\mu}(x_{i})+\delta_{i}(1+k_{i})\{I(y_{i}<c)-\hat{\mu}(x_{i})\}],
\]
\[
\hat{\eta}_{\nni}^{(k)}=\sum_{i=1}^{n}\omega_{i}^{(k)}[\hat{\mu}_{\eta}(x_{i})+\delta_{i}(1+k_{i})\{I(y_{i}<c)-\hat{\mu}_{\eta}(x_{i})\}],
\]
\[
\hat{\xi}_{\nni}^{(k)}(\hat{\xi}_{\nni})=\hat{f}(\hat{\xi}_{\nni})^{-2}\sum_{i=1}^{n}\omega_{i}^{(k)}[\hat{\mu}_{s}(\hat{\xi}_{\nni},x_{i})+\delta_{i}(1+k_{i})\{I(y_{i}\leq\hat{\xi}_{\nni})-\hat{\mu}_{s}(\hat{\xi}_{\nni},x_{i})\}],
\]
where $\hat{\mu}_{\eta}(x)$, $\hat{\mu}_{s}(\xi,x)$ and $\hat{f}(x)$
are nonparametric estimators of $\mu_{\eta}(x)=\pr(y<c\mid x)$, $\mu_{s}(\xi,x)=\pr(y<\xi\mid x)$
and $f(\xi)$, respectively, and $k_{i}$ is the number of times that
$y_{i}$ is selected to impute the missing values of $y$ based on
the original data. These are obtained by kernel regression using a
Gaussian kernel with bandwidth $h=1.5n^{-1/5}$. The variance estimators
are compared in terms of empirical coverage rate and relative bias,
$\{E(\hat{V}_{I})-V\}/V$, where $V$ is the true variance simulated
by Monte Carlo.

Tables \ref{tab:Sim1} and \ref{tab:Sim2} present the simulation
results under simple random sampling and probability proportional
to size sampling, respectively, based on $2,000$ Monte Carlo samples.
Under both sampling designs, the nearest neighbor imputation estimator
has small biases for all parameters $\mu,$ $\eta$ and $\xi$, under
(P1)\textendash (P3) with $m(x)$ accurate approximation for the mean
function and (P4)\textendash (P6) with $m(x)$ inaccurate approximation
of the mean function. For variance estimation, as expected, the naive
jackknife variance estimator is severely biased, indicating that the
lack of smoothness of the matching estimator needs to be taken into
account in variance estimation. In contrast, the proposed jackknife
variance estimators provide satisfactory results under both sampling
designs and for all parameters. The relative biases are small and
the empirical coverage rates are close to the nominal coverage. Overall,
the simulation results suggest that the proposed variance estimator
works reasonably well under the settings we considered. 

\begin{table}
\begin{centering}
{\scriptsize{}{}\caption{{\scriptsize{}{}\label{tab:Sim1}}Simulation results for the population
mean $\mu$, the population proportion $\eta=0.8$ and the population
median $\xi$ under simple random sampling: Bias ($\times10^{2}$)
and S.E. ($\times10^{2}$) of the point estimator, Relative Bias of
jackknife variance estimates ($\times10^{2}$) and Coverage Rate ($\%$)
of $95\%$ confidence intervals.}
} 
\par\end{centering}

\begin{centering}
\begin{tabular}{ccccccccc}
\hline 
\multicolumn{9}{c}{Simple Random Sampling }\tabularnewline
\hline 
 &  &  &  &  & \multicolumn{2}{c}{Prop JK} & \multicolumn{2}{c}{Naive JK}\tabularnewline
 &  & $m(x)$ & Bias  & S.E.  & RB  & CR  & RB  & CR \tabularnewline
\hline 
 & (P1)  & a & 0.00 & 4.87 & 0.1 & 94.9 & \textgreater{}1000 & 100\tabularnewline
$\mu$  & (P2)  & a & 0.12  & 6.08 & 0.5 & 95.3 & \textgreater{}1000 & 100\tabularnewline
 & (P3)  & a & 1.09 & 8.42 & 2.2  & 95.3 & \textgreater{}1000 & 100\tabularnewline
 & (P4)  & i & -0.10 & 5.41  & 3.6  & 96.0 & \textgreater{}1000 & 100\tabularnewline
 & (P5)  & i & 0.20  & 6.59  & 0.1 & 95.4 & \textgreater{}1000 & 100\tabularnewline
 & (P6)  & i & 1.17 & 8.81 & 0.3  & 94.8 & \textgreater{}1000 & 100\tabularnewline
\hline 
 & (P1)  & a & 0.00 & 1.77 & 0.4 & 95.0  & \textgreater{}1000 & 100\tabularnewline
$\eta$  & (P2)  & a & 0.00 & 1.53 & -0.1 & 94.9  & \textgreater{}1000 & 100\tabularnewline
 & (P3)  & a & -0.01 & 1.50 & -5.1 & 94.7 & \textgreater{}1000 & 100\tabularnewline
 & (P4)  & i & 0.03 & 1.63  & 6.1 & 95.4  & \textgreater{}1000 & 100\tabularnewline
 & (P5)  & i & 0.05 & 1.48 & 4.3 & 95.5 & \textgreater{}1000 & 100\tabularnewline
 & (P6)  & i & -0.01 & 1.47 & -0.7  & 94.9 & \textgreater{}1000 & 100\tabularnewline
\hline 
 & (P1)  & a & -0.25 & 6.15 & 2.7 & 94.8 & \textgreater{}1000 & 100\tabularnewline
$\xi$  & (P2)  & a & -0.40 & 7.60 & 2.5 & 94.7 & \textgreater{}1000 & 100\tabularnewline
 & (P3)  & a & -0.37  & 10.19 & 4.0 & 94.6 & \textgreater{}1000 & 100\tabularnewline
 & (P4)  & i & -0.25  & 7.09 & 3.2 & 94.6 & \textgreater{}1000 & 100\tabularnewline
 & (P5)  & i & -0.35 & 8.17 & 7.2 & 96.0 & \textgreater{}1000 & 100\tabularnewline
 & (P6)  & i & -0.54 & 10.78 & 1.8 & 94.1 & \textgreater{}1000 & 100\tabularnewline
\hline 
\end{tabular}
\par\end{centering}
Prop JK: proposed jackknife variance estimation; Naive JK: naive jackknife
variance estimation. a: accurate and i: inaccurate. 
\end{table}
\begin{table}
\begin{centering}
{\scriptsize{}{}\caption{{\scriptsize{}{}\label{tab:Sim2}}Simulation results for the population
mean $\mu$, the population proportion $\eta=0.8$ and the population
median $\xi$ under probability proportional to size sampling: Bias
($\times10^{2}$) and S.E. ($\times10^{2}$) of the point estimator,
Relative Bias of jackknife variance estimates ($\times10^{2}$) and
Coverage Rate ($\%$) of $95\%$ confidence intervals.}
} 
\par\end{centering}
\begin{centering}
\begin{tabular}{ccccccccc}
\hline 
\multicolumn{9}{c}{Probability Proportional to Size }\tabularnewline
\hline 
 &  &  &  &  & \multicolumn{2}{c}{Prop JK} & \multicolumn{2}{c}{Naive JK}\tabularnewline
 &  & $m(x)$ & Bias  & S.E.  & RB  & CR  & RB  & CR \tabularnewline
\hline 
 & (P1)  & a & 0.07 & 4.71 & 1.8 & 95.4 & \textgreater{}1000 & 100\tabularnewline
$\mu$  & (P2)  & a & 0.20 & 5.71 & 6.1 & 95.9 & \textgreater{}1000 & 100\tabularnewline
 & (P3)  & a & 0.73 & 7.71 & 6.0 & 96.1  & \textgreater{}1000 & 100\tabularnewline
 & (P4)  & i & -0.06 & 5.29 & 2.4  & 95.5 & \textgreater{}1000 & 100\tabularnewline
 & (P5)  & i & 0.22 & 6.08 & 7.0 & 95.9 & \textgreater{}1000 & 100\tabularnewline
 & (P6)  & i & 0.99 & 8.23 & 5.4 & 95.1 & \textgreater{}1000 & 100\tabularnewline
\hline 
 & (P1)  & a & -0.01 & 1.89 & -6.0 & 94.5 & \textgreater{}1000 & 100\tabularnewline
$\eta$  & (P2)  & a & 0.02 & 1.63 & -1.9 & 95.3 & \textgreater{}1000 & 100\tabularnewline
 & (P3)  & a & 0.08 & 1.66 & -5.5 & 94.4 & \textgreater{}1000 & 100\tabularnewline
 & (P4)  & i & 0.02 & 1.79 & -4.0 & 95.2 & \textgreater{}1000 & 100\tabularnewline
 & (P5)  & i & 0.03 & 1.60 & 1.8 & 95.2 & \textgreater{}1000 & 100\tabularnewline
 & (P6)  & i & 0.08 & 1.67 & -8.7 & 93.7 & \textgreater{}1000 & 100\tabularnewline
\hline 
 & (P1)  & a & -0.31 & 6.34 & 6.2 & 94.8 & \textgreater{}1000 & 100\tabularnewline
$\xi$  & (P2)  & a & -0.06 & 8.30 & 0.8 & 94.5 & \textgreater{}1000 & 100\tabularnewline
 & (P3)  & a & -0.42 & 11.36 & 5.4 & 94.6  & \textgreater{}1000 & 100\tabularnewline
 & (P4)  & i & -0.32 & 7.57 & 4.1 & 94.0 & \textgreater{}1000 & 100\tabularnewline
 & (P5)  & i & -0.34 & 8.91 & 7.0 & 94.8 & \textgreater{}1000 & 100\tabularnewline
 & (P6)  & i & -0.49 & 12.22 & 2.2  & 94.4  & \textgreater{}1000 & 100\tabularnewline
\hline 
\end{tabular}
\par\end{centering}
Prop JK: proposed jackknife variance estimation; Naive JK: naive jackknife
variance estimation. a: accurate and i: inaccurate. 
\end{table}

\section{Discussion}

Instead of choosing the nearest neighbor as a donor for missing items,
we can consider fractional imputation \citep{kim2004fractional,yang2016fi}
using $K$ $(K>1)$ nearest neighbors. Such extension remains an interesting
avenue for future research.

\section*{Appendix }

\global\long\def\theequation{A\arabic{equation}}
 \setcounter{equation}{0}

\global\long\def\thesection{A\arabic{section}}
 \setcounter{equation}{0}

\global\long\def\thetable{A\arabic{table}}
 \setcounter{equation}{0}

\global\long\def\theexample{A\arabic{example}}
 \setcounter{equation}{0}

\global\long\def\thetheorem{A\arabic{theorem}}
 \setcounter{equation}{0}

\global\long\def\thecondition{A\arabic{condition}}
 \setcounter{equation}{0}

\global\long\def\theremark{A\arabic{remark}}
 \setcounter{equation}{0}

\global\long\def\thestep{A\arabic{step}}
 \setcounter{equation}{0}

\global\long\def\theassumption{A\arabic{assumption}}
 \setcounter{equation}{0}

\global\long\def\theproof{A\arabic{proof}}
 \setcounter{equation}{0}

The Appendix includes proofs of Theorems 1 and 2 and additional assumptions. 

\section{Proof for Theorem 1}

With a scalar matching variable $m$, we have
\begin{eqnarray*}
B_{N} & = & \frac{n^{1/2}}{N}\sum_{i\in A}\frac{1}{\pi_{i}}(1-\delta_{i})\{\mu_{g}(x_{i(1)})-\mu_{g}(x_{i})\}\\
 & \leq & \frac{n^{1/2}}{N}\sum_{i\in A}\frac{1}{\pi_{i}}(1-\delta_{i})\mid m_{i(1)}-m_{i}\mid=o_{p}(1),
\end{eqnarray*}
where $\leq$ in the second line follows by Assumption \ref{asmp:m}
(ii). Based on the decomposition in (\ref{eq:decomposition}), we
can write 
\begin{equation}
n^{1/2}(\hat{\mu}_{g,\nni}-\mu_{g})=D_{N}+o_{p}(1),\label{eq:A1}
\end{equation}
where $D_{N}$ is defined in (\ref{eq:DN}). Then, to study the asymptotic
properties of $n^{1/2}(\hat{\mu}_{g,\nni}-\mu_{g})$, we only need
to study the asymptotic properties of $D_{N}$. For simplicity, we
introduce the following notation: $\mu_{g,i}=\mu_{g}(x_{i})\equiv E\{g(y)\mid x_{i}\}$
and $e_{i}=g(y_{i})-\mu_{g,i}$. We express 
\begin{eqnarray}
D_{N} & = & \frac{n^{1/2}}{N}\left[\sum_{i\in A}\frac{1}{\pi_{i}}\left\{ \mu_{g,i}+\delta_{i}(1+k_{i})e_{i}\right\} -\sum_{i=1}^{N}g(y_{i})\right]\nonumber \\
 & = & \frac{n^{1/2}}{N}\sum_{i=1}^{N}\left(\frac{I_{i}}{\pi_{i}}-1\right)\mu_{g,i}+\frac{n^{1/2}}{N}\sum_{i=1}^{N}\left\{ \frac{I_{i}}{\pi_{i}}\delta_{i}(1+k_{i})-1\right\} e_{i},\label{eq:A2}
\end{eqnarray}
and we can verify that the covariance of the two terms in (\ref{eq:A2})
is zero. Thus, 
\[
\var(D_{N})=\var\left\{ \frac{n^{1/2}}{N}\sum_{i=1}^{N}\left(\frac{I_{i}}{\pi_{i}}-1\right)\mu_{g,i}\right\} +\var\left[\frac{n^{1/2}}{N}\sum_{i=1}^{N}\left\{ \frac{I_{i}}{\pi_{i}}\delta_{i}(1+k_{i})-1\right\} e_{i}\right].
\]
The first term, as $n\rightarrow\infty$, becomes 
\[
V_{g}^{\mu}=\lim_{n\rightarrow\infty}\frac{n}{N^{2}}E\left\{ \var_{p}\left(\sum_{i\in A}\frac{\mu_{g,i}}{\pi_{i}}\right)\right\} ,
\]
and the second term, as $n\rightarrow\infty$, becomes 
\[
V_{g}^{e}=\plim\frac{n}{N^{2}}\sum_{i=1}^{N}\left\{ \frac{I_{i}}{\pi_{i}}\delta_{i}(1+k_{i})-1\right\} ^{2}\var(e_{i}\mid x_{i}).
\]
The remaining is to show that $V_{g}^{e}=O(1)$. To do this, the key
is to show that the moments of $k_{i}$ are bounded. Under Assumption
\ref{asmp:sampling}, it is easy to verify that 
\begin{equation}
\underbar{\ensuremath{\omega}}\tilde{k}_{i}\leq k_{i}\leq\bar{\omega}\tilde{k}_{i},\label{eq:A3}
\end{equation}
for some constants $\underbar{\ensuremath{\omega}}$ and $\bar{\omega}$,
where $\tilde{k}_{i}=\sum_{j=1}^{n}(1-\delta_{j})d_{ij}$ is the number
of unit $i$ used as a match for the nonrespondents. Under Assumption
\ref{asmp:m}, $\tilde{k}_{i}=O_{p}(1)$ and $E(\tilde{k}_{i})$ and
$E(\tilde{k}_{i}^{2})$ are uniformly bounded over $n$ (\citealp{abadie2006large},
Lemma 3); therefore, together with (\ref{eq:A3}), we have $k_{i}=O_{p}(1)$
and $E(k_{i})$ and $E(k_{i}^{2})$ are uniformly bounded over $n$.
Therefore, a simple algebra yields $V_{g}^{e}=O(1)$.

Combining all results, the asymptotic variance of $n^{1/2}(\hat{\mu}_{g,\nni}-\mu_{g})$
is $V_{g}^{\mu}+V_{g}^{e}$. By the central limit theorem, the result
in Theorem \ref{Thm:1} follows.

\section{Proof for Theorem 2}

We impose the following assumptions for the population parameter $\xi_{N}$
and the population estimating function $S_{N}(\cdot)$; see also \citet{wang2011asymptotic}.

\begin{assumption}\label{asmp:sN} 
\begin{description}
\item [{(i)}] The population parameter $\xi_{N}$ lies in a closed interval
$\It_{\xi}$ on $\mathcal{R}$; 
\item [{(ii)}] the function $s(\cdot)$ is bounded; 
\item [{(iii)}] the population estimating function $S_{N}(\xi)$ converges
to $S(\xi)$ uniformly on $\It_{\xi}$ as $N\rightarrow\infty$, and
the equation $S(\xi)=0$ has a unique root in the interior of $\It_{\xi}$;
\item [{(iv)}] the limiting function $S(\xi)$ is strictly increasing and
absolutely continuous with finite first derivative in $\It_{\xi}$,
and the derivative $S'(\xi)$ is bounded away from $0$ for $\xi$
in $\It_{\xi}$;
\item [{(v)}] the population quantities 
\[
\sup_{\xi\in\It_{s}}N^{\alpha}|S_{N}(\xi_{N}+N^{-\alpha}\xi)-S_{N}(\xi_{N})-S(\xi_{N}+N^{-\alpha}\xi)-S(\xi_{N})|\rightarrow0,
\]
and 
\[
\sup_{\xi\in\It_{s}}N^{-1}\sum_{i=1}^{N}|s(y_{i}-\xi_{N}-N^{-\alpha}\xi)-s(y_{i}-\xi_{N})|=O_{p}(N^{-\alpha}),
\]
where $\It_{s}$ is a large enough compact set in $\mathcal{R}$ and
$\alpha\in(1/4,1/2]$. 
\end{description}
\end{assumption}

Assumption \ref{asmp:sN} (v) holds with probability one under suitable
assumptions on the probability mechanism generating the $y_{i}$'s
and on the function $s(\cdot)$, and therefore is justifiable. Under
Assumption \ref{asmp:sN}, by the standard arguments from the theory
on M-estimators \citep{serfling1980approximation}, $\hat{\xi}_{\nni}$
is consistent for $\xi_{N}$. We further make the following assumption.

\begin{assumption}\label{asmp:sN-1} The nearest neighbor imputation
estimator $\hat{\xi}_{\nni}$ is root-$n$ consistent for $\xi_{N},$

\end{assumption}

Now, we give proof for Theorem \ref{Thm:2}. Under Assumptions \ref{asmp:sN}
and \ref{asmp:sN-1}, we can write
\begin{equation}
\hat{S}_{\nni}(\hat{\xi}_{\nni})-S_{N}(\xi_{N})=\{\hat{S}_{\nni}(\xi_{N})-S_{N}(\xi_{N})\}+S'(\xi_{N})(\hat{\xi}_{\nni}-\xi_{N})+o_{p}(n^{-1/2}).\label{eq:s1}
\end{equation}
By Assumption \ref{asmp:sN} (iv), $S(\xi)$ is smooth, and therefore
$S_{N}(\xi_{N})=O_{p}(N^{-1})$, $\hat{S}_{\nni}(\hat{\xi}_{\nni})=O_{p}(n^{-1})$,
and the left hand side of (\ref{eq:s1}) is $o_{p}(n^{-1/2})$. Therefore,
we can obtain a linearization for $\hat{\xi}_{\nni}$ as in (\ref{eq:s2}).

Based on the linearization (\ref{eq:s2}), the asymptotic variance
$V_{\xi}=\dot{S}(\xi)^{-2}\var\{\hat{S}_{\nni}(\xi)\}$. Following
a similar derivation in the proof for Theorem 1, it is easy to show
that
\begin{multline*}
\var\{\hat{S}_{N}(\xi)\}=\lim_{n\rightarrow\infty}\frac{n}{N^{2}}E\left(\var_{p}\left[\sum_{i\in A}\frac{E\{s(y_{i}-\xi)\mid x_{i}\}}{\pi_{i}}\right]\right)\\
+\plim\frac{n}{N^{2}}\sum_{i=1}^{N}\left\{ \frac{I_{i}}{\pi_{i}}\delta_{i}(1+k_{i})-1\right\} ^{2}\var\left[s(y_{i}-\xi)-E\{s(y_{i}-\xi)\mid x_{i}\}\mid x_{i}\right].
\end{multline*}

\section{Assumptions}

\begin{assumption}\label{asump:10}The following conditions hold
for kernel function $K(\cdot)$ and bandwidth $h$:
\begin{description}
\item [{(i)}] the kernel function $K(\cdot)$ is absolutely continuous
with nonzero finite derivative $K'(\cdot)$ and $\int K(x)\de x=1$;
\item [{(ii)}] the bandwidth $h\rightarrow0$ and $nh\rightarrow\infty$
as $n\rightarrow\infty$; 
\item [{(iii)}] there exists a constant $c$, such that $|h^{-1}K'(x_{1}/h)-h^{-1}K'(x_{2}/h)|\leq c|x_{1}-x_{2}|$
for any $x_{1}$, $x_{2}$ and $h$ arbitrarily small. 
\end{description}
\end{assumption} 

Assumption \ref{asump:10} states conditions on the smoothness and
tail behavior of the kernel functions. Popular kernel functions, including
Epanechnikov, Gaussian, and triangle kernels, satisfy the required
conditions.

\bibliographystyle{dcu}
\bibliography{pfi_MIsurvey_v6}

\end{document}